\documentclass[10pt]{article}

\usepackage{amsmath}
\usepackage{amssymb}

\usepackage{graphicx}

\usepackage{cite}

\usepackage{color} 

\usepackage[labelfont=bf,labelsep=period,justification=raggedright]{caption}

\makeatletter
\renewcommand{\@biblabel}[1]{\quad#1.}
\makeatother

\date{}

\begin{document}

% Title must be 150 characters or less
\begin{flushleft}
{\Large
\textbf{Tracking Traders' Understanding of the Market Using e-Communication Data}
}
% Insert Author names, affiliations and corresponding author email.
\\
Serguei Saavedra$^{1,2,3}$, 
Jordi Duch$^{4}$, 
Brian Uzzi$^{1,2,\ast}$
\\
\bf{1} Northwestern Institute on Complex Systems, Northwestern University, Evanston, Illinois, USA
\\
\bf{2} Kellogg School of Management, Northwestern University, Evanston, Illinois, USA
\\
\bf{3} Northwestern University Clinical and Translational Sciences Institute, Northwestern University, Chicago, Illinois
\\
\bf{4} Department of Computer Science and Mathematics, Universitat Rovira i Virgili, Tarragona, Spain
\\
$\ast$ E-mail: uzzi@northwestern.edu
\end{flushleft}

% Please keep the abstract between 250 and 300 words
\section*{Abstract}
Tracking the volume of keywords in Internet searches, message boards, or Tweets has provided an alternative for following or predicting associations between popular interest or disease incidences. Here, we extend that research by examining the role of e-communications among day traders and their collective understanding of the market. Our study introduces a general method that focuses on bundles of words that behave differently from daily communication routines, and uses original data covering the content of instant messages among all day traders at a trading firm over a 40-month period. Analyses show that two word bundles convey traders' understanding of same day market events and potential next day market events. We find that when market volatility is high, traders' communications are dominated by same day events, and when volatility is low, communications are dominated by next day events. We show that the stronger the traders' attention to either same day or next day events, the higher their collective trading performance. We conclude that e-communication among traders is a product of mass collaboration over diverse viewpoints that embodies unique information about their weak or strong understanding of the market.

\section*{Introduction}
Sir Francis Galton's {\it vox populi} conjecture \cite{Galton} that the average estimate of many individuals can exceed individual wit has grown in promise as complex systems become more intricate, interrelated, and immense. Sir Galton's insight laid a foundation for the idea of ``collective wisdom" and represents an emerging interdisciplinary study of how collective information can be leveraged to increase our understanding of large-scale social and economic events \cite{Page,Watts,Lazer,Vespignani}. For example, research embracing the promise of widely available Internet-based data finds that shifts in the volume of keywords in Google searches or Tweets, can detect flu rates, public moods, and consumer demand and prices \cite{Ginsberg,Preis,Goel,Oconnor,Bollen}. This research benefits from the existence of preselected, recognizable words that reflect popular interest or sentiment levels---like the name of a movie or an infectious disease. However, a population's understanding of large-scale phenomena emerges in large part through social collaboration, learning and reasoning, not just interest level \cite{Watts,Lazer,Vespignani,Saavedra10,Page,Barabasi}. Similarly, it has been shown that words derive meaning from the simultaneous association with other words driven by how people characterize and respond to the world around them \cite{Pinker,Sole}. This suggests that the social dynamics captured by bundles of unique and correlated words can summarize the dynamics of single titles and provide relevant information about a population's understanding of complex systems.

In this paper, we present and test a method for capturing the collective understanding of socioeconomic events using e-communication data by inductively identifying bundles of words that significantly deviate from daily communication routines. The rationale of this method is that non-routinary words, whose daily frequency is not a simple product of the total volume of words, could reveal information external to the communication system \cite{Pinker}. Because the method is not dependent on preselected keywords, it aims to be generalizable. To illustrate our method, we study volatility, a multidimensional construct critical in many complex systems \cite{Scheffer09}. For instance, asthma attacks, epilepsy, or climate shift display valuable precursors characterized by a slowing or quickening of fluctuations in parameter values \cite{Scheffer09}. In politics, it relates to legislation, corruption and civil unrest, and in disease control it is linked to new infection rates \cite{Diermeier,Oconnor,Vespignani,Lazer}. In markets, volatility notably affects all investment decisions \cite{Whaley} and scale dynamics including critical transitions such as financial crashes \cite{Poon,Scheffer09,May}.

In particular, we analyzed the association of traders' person to person communications with their understanding of market volatility. The data we use to identify the collective understanding of the market is distinctive. Unlike past research that has used general public information in Google or Twitter, we draw on the content of instant messages (IMs), an increasingly pervasive form of e-communication \cite{Harris,Antweiler,Stein}. Our data includes the full population of more than 3 million IMs sent and received by all the day traders at a typical trading firm from 1/2007-4/2010. IMs represent an excellent source of traders' collective thinking about the market \cite{Saavedra,Zhao}. Unlike investors who make money by holding stocks that rise in value over the long term, day traders make money by buying and selling many stocks over a single day with regard to movements of their stock prices. Consequently, day traders face the challenge of continually understanding and deciphering how news is affecting, and will affect, market volatility during the day and the next day, their trading horizon. For example, when a news report states a nuclear reactor may fail the ramifications of that news for the market are unknown at the time of the report. Will oil prices rise and by how much? Will nuclear stocks fall with fears of a meltdown or rise with near term oil shortages? For day traders, the answer to these and other questions are solved in large part through informal consultation with their instant message contacts who are doing likewise with their contact network \cite{Saavedra,Kirman}. This communication pattern spreads the IM network of the traders over diverse viewpoints and a broad spectrum of the market (for example, the trader population in our firm trades over 4000 stocks \cite{Saavedra}).

Communication of the above type has been shown to effectively capture the collective knowledge of decision makers, while at the same time, canceling out their individual biases \cite{Page}. And because communication is costly, it is likely that traders exchange IMs that contain groups of words that efficiently convey their understanding of the market \cite{Pinker}. This social dynamic suggests that as traders use a word bundle more than an alternative bundle, an assimilation of thought averaged over the diverse views of many traders has emerged in such a way that an increase in understanding of the market may be embodied in their communications.   

\section*{Results}

\subsection*{Extraction method}
To extract significant information from traders' IMs, we adapted fluctuation scaling techniques \cite{Zoltan,deMenezes}. Step one filtered the population of words to those words appearing $>1000$ times or roughly $>1$ time daily in order to remove misspellings and to consider commonly used words by the majority of traders \cite{Altmann}. Consistent with universal patterns of human language\cite{Zipf}, words in our filtered IM corpus (over 11 million total words and over 232 thousand unique words) appear approximately twice as often as the next least frequent word (Fig. 1A). Step two classified the population of words in our filtered IM corpus into words that follow either the routinary or external factors of the communication system \cite{deMenezes} (Methods). Operationally, words that follow routinary factors have a daily frequency proportional to the total number of daily words (Fig. 1B), suggesting that they are a function of traders' communication routines rather than an exceptional stimulus. Consistent with linguistic research 302 out of 319 English ``stop words" (e.g. a, an, for, or, the) \cite{Luhn}, which are commonly filtered words in text analysis \cite{Luhn,Altmann}, were classified in this category. By contrast, the daily frequency of words following external factors were statistically unrelated to the density of total daily words, suggesting that traders use these words to characterize external stimuli. This subset of words was defined as extracted words in our analysis. A total of 459 words were extracted. Importantly, Figure 2 shows that extracted words can have different temporal dynamics, revealing that each word characterizes a piece of information from the overall communications among traders. This suggests that bundles of words may provide a more general understanding of the market. 

Step three found bundles of extracted words that were significantly correlated with each other and weakly correlated with other extracted words based on their daily pairwise frequency. For each pair of extracted words $i$ and $j$, we calculated the Pearson pairwise correlation $\rho_{ij}(\Delta_{fi},\Delta_{fj})$, where $\Delta_f$ is the vector of frequency changes. To appropriately quantify the statistical similarity of each pair of words, we compared the observed daily pairwise correlation to a null model where the word pairs were randomly shuffled.  We calculated the expected correlation $\rho^*$ and standard deviation $\sigma(\rho^*)$ from the random model to compute a $z$-score of the observed relative to the random given by $z_{ij}(\rho_{ij})=(\rho_{ij}-\rho_{ij}^*)/\sigma(\rho_{ij}^*)$. Word bundles were then created using a version \cite{Gomez} of the Extremal Optimization Algorithm \cite{Bak,Duch} for community detection in correlation networks. We used words as nodes and the size of $z$-score between words as edge weights to form the correlation network. The number of bundles is not fixed in advance, bundles of words are formed by maximizing the network's modularity parameter \cite{Newman}. This method clusters words that are highly correlated between each other and weakly correlated with a different group of words.   

As a robustness check on the original partition, we performed a second optimization of the modularity parameter based on the Kernighan-Lin algorithm \cite{Newman}. This consists in a fine tuning of the clusters of the original partition with a bootstrapping process \cite{Gomez}. Additionally, we obtained the same number of bundles when we applied a simulated annealing approach to maximize the modularity parameter \cite{Guimera}. Moreover, the partition did not change whether we used the entire dataset or split it into datasets of equal size.

\subsection*{Extracted word bundles}

Three clusters or word bundles were found. Bundles one and two contained $35\%$ and $45\%$ of our extracted words respectively, and were made up of virtually all English words. Bundle three was made up of principally foreign language words, which suggests a connection to a subset of multilingual traders specific to the population characteristics of this trading company. Bundle one contained extracted words such as negative, lows, cuts, insane, crazy, ugly, banks, oil, weak, interest, s\&p. Illustrative examples of bundle two keywords are happy, alert, dollars, excited, bloomberg, reuters, win, trend, china, nyse; while examples of bundle three are prosto, nego, nada, csak, nem. Since there is no a priori reason to expect the resulting grouping of words, any relationship among these words should be treated as a consequence of their own frequency dynamics. Words within a bundle were highly correlated. The proportion of significant correlations ($z$-score$>2$) within bundles was $54\%$, while the proportion between bundles was only $22\%$. These findings confirm that word bundles capture information embedded in words that gain meaning through their co-occurrence \cite{Sole,Pinker}.

\subsection*{Collective understanding} 

To study how traders' communications express their collective understanding of market volatility, we used the daily closing value of market volatility and the daily frequency of word bundles relative to the extracted words. Volatility can be operationalized by the volatility index (VIX) \cite{Whaley,Scheffer09}, which corresponds to the expected future volatility over the next 30 calendar days. The VIX, also known as the ``fear" index, gives a good approximation to the overall sentiment of traders by reflecting the price of portfolio insurance, i.e. the higher the level of uncertainty in the market, the higher the VIX. We measured the relative frequency of each word bundle $i$ as $\gamma_i(t)=\frac{\sum_j^c f_j(t)}{\sum_k^W f_k(t)}$, where $f_j(t)$ is the frequency of word $j$ in day $t$, $c$ is the total number of words in bundle $i$, and $W$ is the total number of extracted words. To analyze the correlation between word bundles and volatility, we transformed all our variables to their first differences \cite{Granger}, $\Theta_i(t)=\gamma_i(t)-\gamma_i(t-1)$. This process also made all our variables stationary (Methods), a characteristic necessary to analyze time series data \cite{Granger}. The cross-correlation $\rho(\Delta t)$ is measured with a time lag parameter $\Delta t$ \cite{Preis} over the day-to-day movements of each word bundle.

For day traders, same day $\Theta_{VIX}(t)$ and next day $\Theta_{VIX}(t+1)$ volatility are critical to understanding the implications of their trading decisions. Figure 3 shows the time series correlations of day-to-day movements of volatility with day-to-day movements of word bundles. A value of $0$ on the x-axis indicates the correlation of same day movements between volatility and the relative frequency of a word bundle, and negative or positive values on the x-axis indicate lead and lag correlations for word bundles respectively. Points above or below the dotted horizontal lines are statistically different from chance. We found that word bundle one was significantly ($p<0.001$) associated with same day movements only (Fig. 3A), suggesting that it captured the collective understanding of same day events $\Theta_{VIX}(t)$. By contrast, word bundle two was significantly ($p<0.001$) associated with next day movements only (Fig. 3B), suggesting that it reflects the collective understanding of potential next day events $\Theta_{VIX}(t+1)$. This predictive utility was confirmed with Granger causality tests \cite{Granger} ($p=0.031$) according to the equation $\Theta_{VIX}(t+1)=\Theta_{VIX}(t)+\Theta_2(t)$. Word bundle three was unrelated to day-to-day movements of any kind, suggesting that the use of foreign language words in the communications in our sample was related to factors other than volatility. Moreover, we found no association whatsoever between the total number of words (Fig. 1) and volatility.

Second, we found systematic associations between the level of volatility and the degree to which same day events dominate traders' communications---what we called temporal understanding. We defined days of low and high volatility by normalizing VIX to a $z$-score using its sample mean and standard deviation. Values of $z_{VIX}(t)>0$ and $z_{VIX}(t)<0$ were defined as days of high and low volatility respectively \cite{Whaley}. We quantified temporal understanding by the degree to which the relative frequency of word bundle one dominated word bundle two each day $C(t)=\gamma_1(t)-\gamma_2(t)$, and normalized that difference with a $z$-score computed on the sample mean and standard deviation. Figure 4 indicates that when the level of volatility was high, the word bundle associated with same day events dominated traders' communications ($z_{C}(t)>0$). Conversely, when volatility was low, the word bundle associated with next day events dominated traders' communications ($z_{C}(t)<0$). The horizontal dashed line is the boundary between the days when one word bundle dominated the other. To guide the eye, the level of temporal understanding changes color to reflect the dominant word bundle. These patterns were confirmed by Fisher's exact test of finding the co-occurrence of these paired events ($p<10^{-12}$, two-sided, see Methods). This result suggests that the collective understanding updated quickly with changes in market conditions. When traders faced high volatility their communications focused on same day events, which is likely an expression of their attempt to reduce the uncertainty presently in the market. By contrast, when traders faced low volatility, their collective understanding turned to the potential next day events, which has relatively more uncertainty for trading prospects.

These empirical regularities point to possibly new relationships between complex system behavior and communications among the participants in the system. Broadly, like past work that has looked at specific preselected keywords, our bundles of words method appears indicate that only a small fraction of the words are used to communicate the system's behavior. However, we find that the essence of communications is not represented by single keywords but by co-occurring words from which collective meaning is mutually constructed. This suggests that while single words may be useful in certain situations, bundles of related words can capture information different from single keywords. Further, we found that separate bundles of words are related to different dimensions of volatility, most importantly the same day and next day, and high and low, volatility in a system.  Finally, the level of collective understanding of same day versus next day events is relative rather than absolute. This suggests that different points of view are simultaneously held by the same population but to different degrees. This raises the interesting and unexpected proposition that the greater the attention to either same day or next day events, the clearer is the collective understanding of the market and vice versa. If this is the case, one would expect that the clearer the understanding of the market, the better their investment decisions, a test we turn to next. 

\subsection*{Collective trading performance}

Finally, we tested if the level of attention to either same day or next day events was associated with the collective trading performance of our population, predicated on the assumption that the greater the attention in word bundles, the greater the collective understanding of the market. This test is novel for our model and the collective wisdom literature which has not examined whether the attention of a group is correlated with the actual collective performance. To capture these dynamics, we measured collective trading performance $p(t)$ as the percentage of traders that made money at the end of the day $t$ in the firm. We operationalized an attention index as the absolute difference between the relative frequencies of word bundle one (same day events) and word bundle two (next day events), $A(t)=|\gamma_1(t)-\gamma_2(t)|$. To appropriately compare these time series, we calculated the correlation between the first differences of collective performance and the first differences of collective attention. First differences are operationalized as the difference between the values at time $t$ and the values at time $t-1$, i.e. $\Theta_{A}(t)=A(t)-A(t-1)$ and $\Theta_{p}(t)=p(t)-p(t-1)$ for collective attention and collective performance respectively.

We found a significant ($p<10^{-4}$), positive correlation of $0.19$ between the first differences of collective performance $\Theta_{p}(t)$ and the first differences of collective attention $\theta_{A}(t)$ (Fig. 5). This positive correlation was supported by the $95\%$ confidence intervals $0.11$-$0.26$ using Fisher's transformation. The statistical significance of this correlation was also confirmed by the lower expected correlation ($0\pm 0.037$) between two independent and identically distributed random variables across the same observation period. Figure 5 shows that as the attention in traders' understanding of same day or next day events increased relative to the previous day, their relative collective performance increased on average. Moreover, when the first differences of underlying volatility (VIX), number of traders and collective attention are added into a regression equation to account for the first differences of collective trading performance, the relationship between collective attention and performance holds. This suggests that traders' attention to events, as captured by word bundles, can reveal traders' collective understanding of the market.
 
\section*{Discussion}
The conjecture that the average collective information of the many is better than the knowledge of any individual has never been more relevant than today, where large-scale social and economic problems such as financial crises or epidemic outbreaks are necessary to anticipate and prevent. While new widely available e-communication data (IMs, email, blogs, message boards) have presented a new opportunity to apply and test Galton's collective wisdom hypothesis, it has also created new challenges. To date, tests have keyed on single preselected words that reflect the intensity of popular interest but increasingly, these data are a co-mingling of many reactions, events, and activities that participants experience simultaneously. We built on this work by offering a method that inductively garners a population's understanding of external events by moving from single preselected keywords to significant behavioral changes in communication routines. Our methodological framework inductively identifies words different from daily communication routines, making it generalizable to other domains and in domain where keywords are unknown a priori. 

Using unique information from more than 3 million IMs sent and received among day traders and their contacts, we showed that just 459 words behave differently from the expected patterns implied by communication routines. Moreover, the 459 words reduced to three word bundles, of which two bundles conveyed traders' understanding of same day and next day market events. When the level of volatility was high, same day events dominated, and when the level of volatility was low the next day events dominated the collective understanding, revealing a link among word bundles, trading horizon and the level of volatility in the market. Importantly, we found that as the level of attention with regard to a specific collective understanding increased, the more the traders appeared to have a clearer understanding of the market, a conclusion supported by their collective trading performance. These results show that non-routinary communication can, in fact, reveal unique information about a populations' understanding of large-scale social and economic dynamics. 

Our work also raises the questions about the micro processes at play that lead to an emergence of collective understanding. While we observe the result of those processes in the form of changes in the frequency of word bundles, we know little about how individuals learn from each other, when and what information solidifies in someone's mind the line between supposition and actionable facts, or even what information attracts attention. Along with this information, tracking how the information may propagate through the IM network can also be valuable in studying the micro foundations of collective understanding. 

\section*{Methods}
{\bf  Ethics Statement}. The study meets all Northwestern University Institutional Review Board (IRB) exemption criteria of anonymity, non-interactivity, and 100\% archival data. Northwestern University IRB stipulates that data that are (1) archival, (2) do not involve interaction with subjects, and (3) are anonymized are IRB exempt. In our case, all three stipulations were met. The data were 100\% archival. The data were 100\% archived before we received it. The data were archived according to well known laws that stipulate that all trading data and all electronic communications of every trader be recorded and stored for 7 years and remain accessible for post trading analysis. Under the same ruling, all the data are considered to be wholly the company's assets. Because these reporting factors are a matter of common knowledge among traders, we sort and received verbal confirmation from the company that all their traders were fully aware of and in voluntary compliance with these record keeping and ownership laws. For example, the company confirmed that all traders at the firm were aware of the legal protocols of trading and that the traders know that 100\% of their electronic communications and trading are recorded by law. We received written approval from the firm to use their data for research purposes and to publish the results of our findings if the name, location, and other defining characteristics of the firm or its traders were kept confidential in accordance with standard research protocols. Also, Northwestern University IRB stipulates that IRB exempt studies must have no interaction with human subjects and that information must be 100\% anonymized. We did not interact with or manipulate human subjects in anyway, all personally identifiable data were 100\% anonymized, and all analyses were conducted on data that had been anonymized using randomized IDs in accordance with the protocols set forth by the firm's information technology officer. The ethic committee was not involved because the data were 100\% archival, had no human subject interaction, and were 100\% anonymized. Research was sponsored by the Army Research Laboratory and was accomplished under Cooperative Agreement Number W911NF-09-2-0053. The views and conclusions contained in this document are those of the authors and should not be interpreted as representing the official policies, either expressed or implied, of the Army Research Laboratory or the U.S. Government. The U.S. Government is authorized to reproduce and distribute reprints for Government purposes notwithstanding any copyright notation here on.  The funders had no role in study design, data collection and analysis, decision to publish, or preparation of the manuscript.
\\ \\
{\bf Extraction method}. We considered extracted words and routinary words, respectively, as words dominated by external and routinary factors of the communication system. For each word $i$, we calculate the strength of its routinary factor by the ratio $\eta_i=\sigma_i^{r}/\sigma_i^{ext}$, where $\sigma_i^{r}$ and $\sigma_i^{ext}$ are the standard deviations of the routinary and external factors respectively. Routinary factors are given by $f_{i}(t)^{r}=<f_i>\frac{N(t)}{<N>}$, i.e. changes in the overall activity of total number of words in day $t$ are reflected in a proportional fashion on the frequency of word $i$ in day $t$. Note that $<f_i>$  and $<N>$ are the average frequency of word $i$ and total daily number of words respectively computed over all activity. Thus, external factors are computed by $f_{i}(t)^{ext}=f_{i}(t)-f_{i}(t)^{r}$, where $f_{i}(t)$ is the observed frequency of word $i$ in the day $t$. Typically, $\eta\gg1$ and $\eta\ll1$  correspond to frequencies dominated by routinary and external factors respectively \cite{deMenezes}. We use a random null hypothesis to appropriately classify words according to their routinary factor strengths. The random null hypothesis is performed by randomly shuffling the daily frequency of each word. For each word $i$, the random null hypothesis is used to compute the expected ratio $\eta_i^*=\sigma_i^{r*}/\sigma_i^{ext*}$ and standard deviation $\sigma(\eta_i^*)$. If words follow routinary factors, we would expect them to behave significantly different from random fluctuations, i.e. following the observed daily communication routine. However, if they follow external factors, we should observe no difference with the random null hypothesis, i.e. words have no correlation with the observed communication routine. Hence, calculating the $z$-score$=(\eta_i-\eta_i^*)/\sigma(\eta_i^*)$, we extracted only words that behave similar to random fluctuations, i.e. $-2<z<2$. We note that words with $\eta<0.35$ follow external factors, which we took as our extracted words.  
\\ \\
{\bf Stationarity}. To ensure that all our data sets (once they were first differenced) were stationary, we used three standard approaches. We cleared the Augmented Dickey-Fuller unit-root test at the $p<10^{-4}$ level, we cleared the Phillips-Perron unit roots test at the $p<10^{-4}$ level, and had negative ($-0.46<\phi<-0.15$) coefficients in all the lagged AR(1) autocorrelation variables.
\\ \\
{\bf Fisher's exact test}. We test the null hypothesis of no association between the two variables $\delta_{C}(t)$ and $\delta_{VIX}(t)$ using Fisher's exact test, where $\delta_{C}(t)$ is defined as a random variable that takes the value of $1$ if traders' communications are dominated by same day events (word bundle one) $z_{C}(t)>0$ at time $t$, and the value of $0$ otherwise. Similarly, we defined $\delta_{VIX}(t)$ as a random variable that takes the value of $1$ if the market is at a state of high volatility $z_{VIX}(t)>0$ at time $t$, and the value of $0$ otherwise. We have 858 business days in the sample, 255 days under high market volatility (198 happened during the dominance of same day events), and 383 days dominated by same day events in total.

\section*{Acknowledgments}
We would like to thank Alex Arenas, Jordi Bascompte, Martha de la Vega, Sergio G\'{o}mez, Roger Guimer\`{a}, Jonathan Haynes, Alejandro Morales Gallardo, Janet Pierrehumbert, Christopher Rhoads and Olivia Woolley for useful discussions that led to the improvement of this work. SS and BU thank the Kellogg School of Management, Northwestern University, the Northwestern Institute on Complex Systems (NICO), and the Army Research Laboratory under Cooperative Agreement W911NF-09-2-0053 for financial support. SS also thanks NUCATS grant UL1RR025741. JD thanks the Spanish Ministry of Science and Technology (FIS2009-13730-C02-02) and the Generalitat de Catalunya (SGR-00838-2009).

%\section*{References}
\bibliography{my_bib}
\clearpage

\section*{Figure Legends}
\begin{figure}[!ht]
\begin{center}
\includegraphics[width=6in]{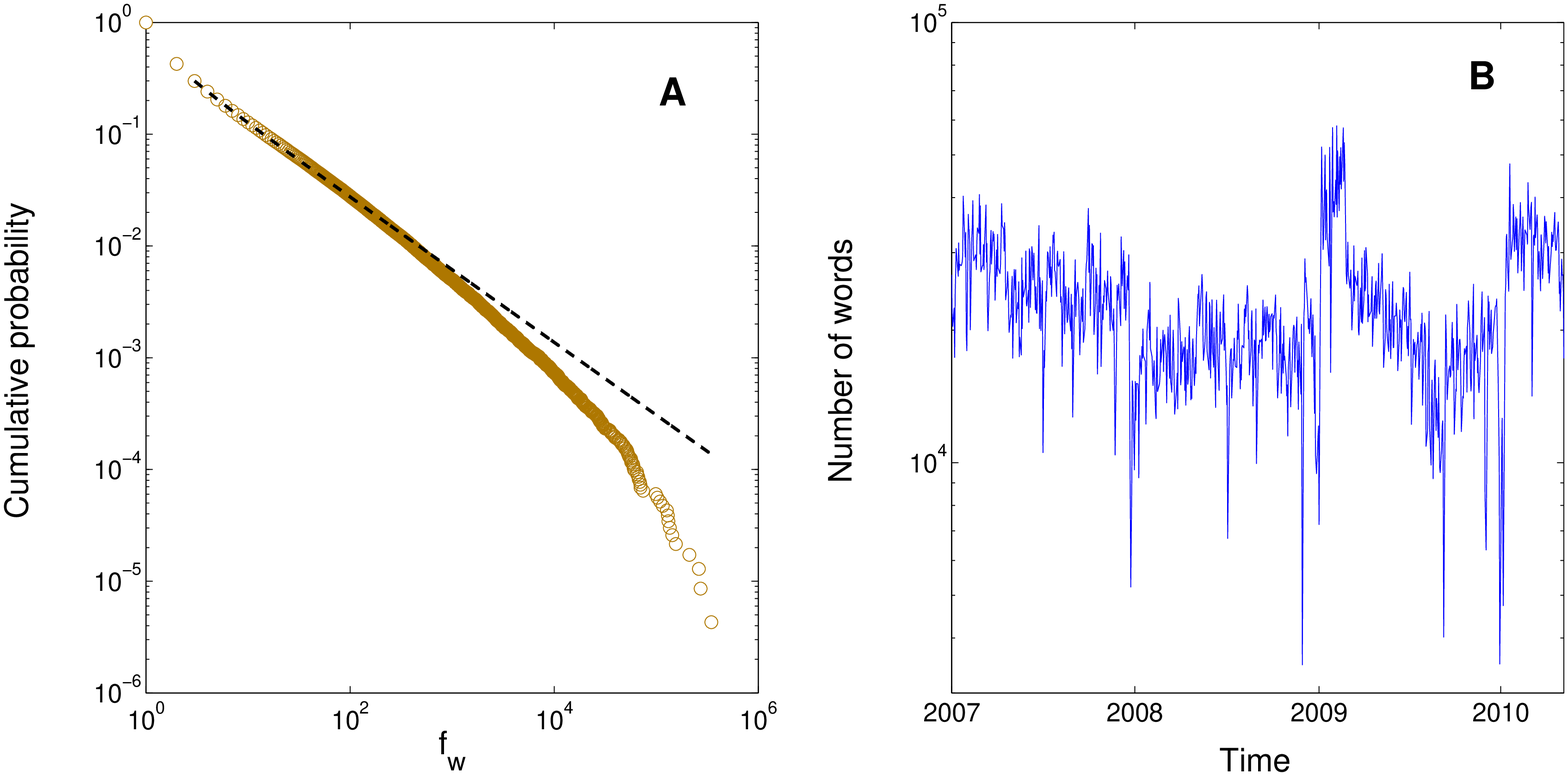}
\end{center}
\caption{
{\bf Communication routines.} Panel {\bf A} shows that the cumulative frequency on a log-log scale of filtered words $f_w$ (viz. word counts) is approximately distributed following Zipf's law according to a power law $P(f_w)\sim f_w^-1.88$ (KS test, $p=0.21$). Panel {\bf B} shows the evolution of traders' messaging volume defined as the total daily number of words on a log scale. Note that the daily frequency of words that follow communication routines can be approximated simply by its global frequency and the total number of words in each day.
}
\label{fig1}
\end{figure}

\begin{figure*}
\begin{center}
\includegraphics*[width=5in]{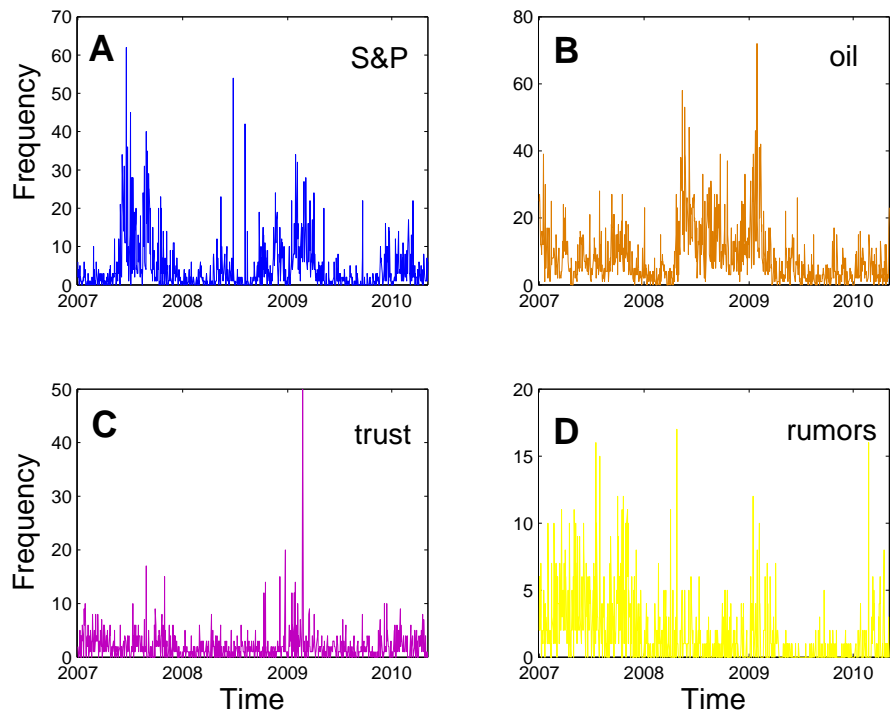}
\end{center}
\caption{Daily frequency of illustrative extracted words. The figure shows the daily frequency (number of times a word is counted each day) of illustrative extracted words {\bf A} S\&P, {\bf B} oil, {\bf C} trust and {\bf D} rumors across the observation period. Note that each word has a unique temporal dynamic characterized by brief periods of intense usage---bursts---preceded by and followed by relatively long periods of low usage.   
}
\label{2}
\end{figure*}

\begin{figure}[!ht]
\begin{center}
\includegraphics[width=6in]{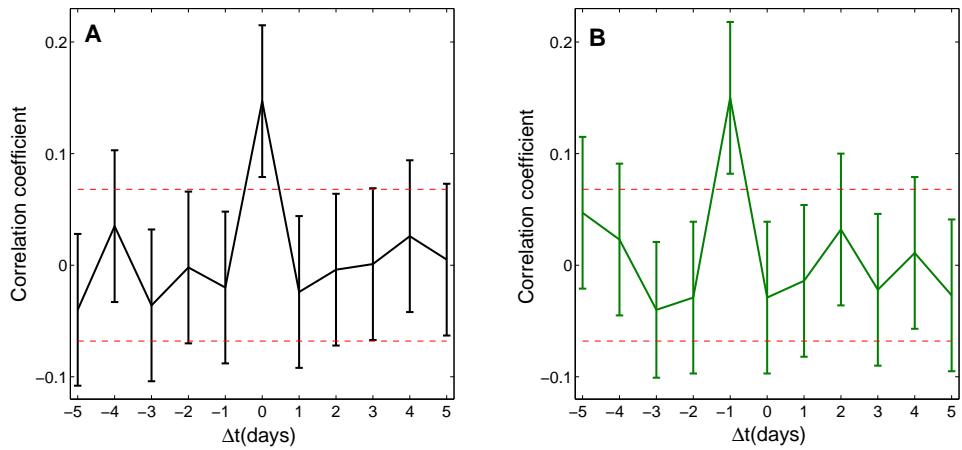}
\end{center}
\caption{
{\bf Cross-correlations between market volatility and word bundles.} The figure shows the time dependent cross-correlations between day-to-day movements of market volatility given by the daily changes in closing values of VIX \cite{Whaley} at time $t$, and {\bf A} the day-to-day movements of relative frequency of word bundle one over $\Delta t$ and {\bf B} word bundle two over $\Delta t$. Day-to-day movements are calculated using the first differences. High-low bars indicate the $95\%$ confidence intervals using Fisher's transformation. The red dashed lines indicate the $95\%$ confidence interval for cross-correlations of two independent and identically distributed random variables across the same observation period.
}
\label{fig3}
\end{figure}

\begin{figure}[!ht]
\begin{center}
\includegraphics[width=3.5in]{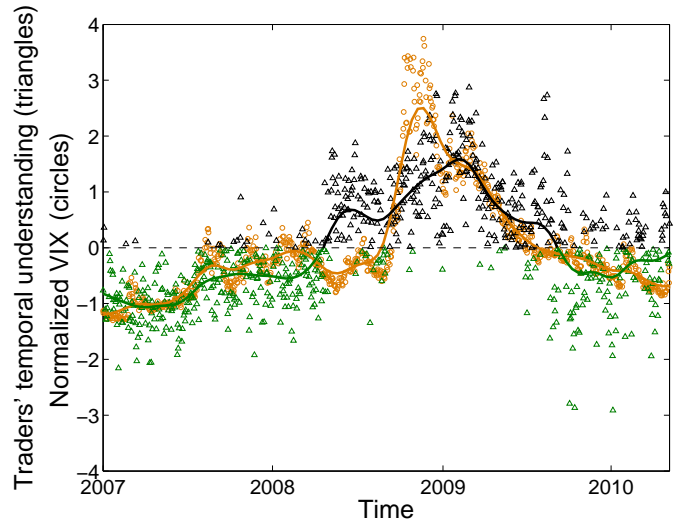}
\end{center}
\caption{
{\bf Traders' temporal understanding of market volatility.}  Triangles show the relative frequency of traders' understanding of same day events (word bundle one) compared to the relative frequency of traders' understanding of next day events (word bundle two) as given by $C(t)=\gamma_1(t)-\gamma_2(t)$ normalized to the $z$-score using the sample mean and standard deviation. The orange circles show market volatility over time normalized to the $z$-score using the sample mean and standard deviation calculated over all the data. Orange circles above dashed line represent days of high and low volatility respectively. Black triangles above the dashed line represent days when word bundle one dominated ($z_{C}(t)>0$), and green triangles below the dashed line represent days when word bundle two dominated the content of traders' IM communications ($z_{C}(t)<0$). These time-series patterns indicate that same day events (word bundle one) systematically dominate traders' understanding on days of high volatility while next day events (word bundle two) systematically dominate on days of low volatility. For visibility purposes, solid lines correspond to the moving average computed using a kernel smoothing regression with a window of one month.
}
\label{fig4}
\end{figure}

\begin{figure}[!ht]
\begin{center}
\includegraphics[width=3in]{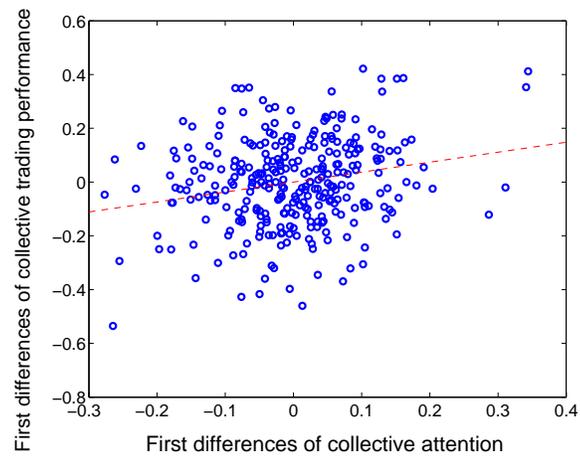}
\end{center}
\caption{ 
{\bf Correlation between collective attention and collective trading performance.} The figure shows the relationship between the first differences of collective attention $\Theta_{A}(t)=A(t)-A(t-1)$ (x-axis) and first differences of collective trading performance $\Theta_{p}(t)=p(t)-p(t-1)$ (y-axis). We found a significant ($p<10^{-4}$), positive correlation of $0.19$. The red dashed line corresponds to the best linear fit ($p=0.002$) over all data points and it is used only to guide the eye for the positive correlation.
}
\label{fig5}
\end{figure}

\end{document}